\documentclass[10pt]{article}

\usepackage{amsmath,amsthm,amsfonts,amscd,eucal,latexsym,amssymb}
\usepackage{epsfig}   
\input xy
\xyoption{all}

\def\phi{\varphi}

\newcommand{\T}{\mathcal{T}}

\newcommand{\Dg}{\mathcal{D}\gamma}

\def\P{\mathfrak{P}}

\def\Laenge{\mathrm{L}}

\newcommand{\E}{\mathrm{E}}

\def\epsilon{\varepsilon}

\def\>{\rangle}

\newsymbol\bt 1202           

\def\bR{{\mathbb R}}

\newsymbol \rest 1316         
 
\def\gC{{\mathfrak C}}

\def\beq{\begin{eqnarray}}
\def\eeq{\end{eqnarray}}

\newcommand{\ca}[1]{{\cal #1}}         

\newcounter{proposition}[section]
\newcounter{theorem}[section]
\newcounter{lemma}[section]
\newcounter{definition}[section]
\newcounter{remark}[section]
\newcounter{conjecture}[section]
\newcounter{corollary}[section]

\def\theproposition{\thesection.\arabic{proposition}}
\def\thetheorem{\thesection.\arabic{theorem}}
\def\thelemma{\thesection.\arabic{lemma}}
\def\thedefinition{\thesection.\arabic{definition}}
\def\theremark{\thesection.\arabic{remark}}
\def\theconjecture{\thesection.\arabic{conjecture}}
\def\thecorollary{\thesection.\arabic{corollary}}

\def\s #1 {\section{#1}}

\def\ssa #1 {\ifhmode{\par}\fi\refstepcounter{subsection}
  \noindent {\bf\thesubsection}. {\em #1}.\quad
  \addcontentsline{toc}{subsection}{\protect\numberline{\thesubsection} #1}%
  }

\def\ssb #1 {\ifhmode{\par}\fi\refstepcounter{subsection}
  \noindent {\bf\thesubsection.} {\em #1.}\quad
  \addcontentsline{toc}{subsection}{\protect\numberline{\thesubsection} #1}%
  }

\def\proposizione {\ifhmode{\par}\fi\refstepcounter{proposition}
  \noindent {\bf Proposition \theproposition}. \quad}
\def\teorema {\ifhmode{\par}\fi\refstepcounter{theorem}
  \noindent {\bf Theorem \thetheorem}. \quad}
\def\lemma {\ifhmode{\par}\fi\refstepcounter{lemma}
  \noindent {\bf Lemma \thelemma}. \quad}
\def\definizione {\ifhmode{\par}\fi\refstepcounter{definition}
  \noindent {\bf Definition \thedefinition}. \quad}
\def\remark {\ifhmode{\par}\fi\refstepcounter{remark}
  \noindent {\bf Remark \theremark}. \quad}
\def\congettura {\ifhmode{\par}\fi\refstepcounter{conjecture}
  \noindent {\bf Conjecture \theconjecture}. \quad}
\def\corollario {\ifhmode{\par}\fi\refstepcounter{corollary}
  \noindent {\bf Corollary \thecorollary}. \quad}

\usepackage{amsfonts}
\usepackage{graphicx}

\begin{document}


\title{Two-dimensional Ricci flow as a stochastic process}


\author{Marco Frasca \\ Via Erasmo Gattamelata, 3 \\ 00176 Roma (Italy)}


\date{\today}



\maketitle

\begin{abstract}
We prove that, for a two-dimensional Riemannian manifold, the Ricci flow is obtained by a Wiener process.
\end{abstract}


\section{Introduction}

A heat equation in $\bR^n$ is obtained through a Brownian motion. One has a Wiener measure $dW(\gamma)=\frac{1}{Z}e^{-\frac{1}{2}E(\gamma)}{\ca D}\gamma$ on the continuos path space $\gC_x(\bR^n,t)$ starting at $\gamma(0)=x$ on the time interval $[0,t]$. So, a kernel $\cal K$ solving the heat equation
\begin{equation}
    \frac{\partial{\cal K}}{\partial t}=\frac{1}{2}\sum_{i=1}^n\frac{\partial^2}{\partial x_i^2}{\cal K}
\end{equation}
for a given initial condition ${\cal K}(x,0)=u(x)$ can be written as the Wiener integral
\begin{equation}
    {\cal K}(x,t)=\frac{1}{Z}\int_{\gC_x(\bR^n,t)}e^{-\frac{1}{2}E(\gamma)}u(\gamma(t)){\ca D}\gamma.
\end{equation}
This kind of solutions are known in physics literature as path integrals and describe a physical process of random motion of a particle in a fluid undergoing casual scattering by the fluid molecules. Quite recently this integral has been generalized to a compact Riemannian manifold by B\"ar and Pf\"affle \cite{BaPf}. These authors obtained the following result:

\begin{teorema}\label{thm:PathIntegral1}
Let $M$ be an $m$-dimensional closed Riemannian manifold, let $E$ be a vector
bundle over $M$ with a metric and a compatible connection $\nabla$.
Let $H = \frac12\nabla^*\nabla + V$ be a self-adjoint generalized Laplace
operator acting on sections in $E$.
Let $t>0$.

Then for any sequence of partitions $\T_n =(t_1^n, \ldots,t_{r_n}^n)$ with
$|\T_n| \to 0$ and $\Laenge(\T_n) \to t$ as $n \to \infty$ and for any $u\in
C^0(M,E)$ 
\begin{equation*}
\frac{1}{Z(\T_n,m)} \,
\int_{\P_x(M,\T_n)} 
\exp\left(-\frac12\E(\gamma)+\int_0^{\Laenge(\T_n)}\frac16 R(\gamma(s))\, 
ds\right)\cdot
\tau(\gamma,\nabla)_{\Laenge(\T_n)}^{0}\times
\end{equation*}
\begin{equation*}
\times
\prod_{j=1}^{r_n}
\exp\left(-\int_{\sigma_{j-1}(\T_n)}^{\sigma_j(\T_n)}
  \tau(\gamma,\nabla)_s^{\Laenge(\T_n)}\cdot
  V(\gamma(s)) 
  \cdot\tau(\gamma,\nabla)_{\Laenge(\T_n)}^{s}\,ds\right)\cdot 
u(\gamma(\Laenge(\T_n)))\,\Dg
\end{equation*}
\begin{equation*}
\quad\quad\xrightarrow{n \to \infty} \quad\quad e^{-tH}u(x)
\end{equation*}
converges uniformly in $x$.
\end{teorema}

For the above result is

\begin{equation}
Z(\T,m) := \prod_{j=1}^r(4\pi t_j)^{m/2} .
\label{eq:ZTerm}  
\end{equation}

The question we ask in this letter is what is the Brownian motion underlying the Ricci flow. We will do this in two dimensions being this the most straightforward result to be obtained. This opens a problem on how to generalize this result to higher dimensions. We just note that the Perelman ${\cal L}$-length functional \cite{MorGan} plays a crucial role here as, when we look at the B\"ar and Pf\"affle theorem, we can recognize this functional in the exponential given by
\begin{equation}
    {\cal L}(\gamma)=\frac12\E(\gamma)-\int_0^{\Laenge(\T_n)}\frac16 R(\gamma(s))\,ds
\end{equation}
for each partition. This result was already obtained in physics by Bryce DeWitt in the context of a formulation of Feynman path integrals for a non-Euclidean manifold \cite{DeWitt}.

\section{Main Theorem}


We prove the following theorem:

\begin{teorema}
\label{teo1} 
Given a Riemann manifold $(M^2,g)$, for a metric starting at $g(x,0)=g_0(x)$, the Ricci flow is given by a Wiener integral.
\end{teorema}

\begin{proof}
Let us consider Ricci flow in harmonic coordinates, $\triangle x_i=0$. One has
\begin{equation}
    \frac{\partial g}{\partial t} = \triangle g + 2Q(g\partial^{-1}g)
\end{equation}
being $Q$ a quadratic form in $g^{-1}$ and $\partial g$ and so, containing only lower order derivatives of $g$. For a two-dimensional Riemannian manifold one can always find a diffeomorphism such that $Q=0$ and the Ricci flow is just the heat equation for the metric. Then, applying theorem \ref{thm:PathIntegral1}, we can write the solution of this equation through a Wiener integral with the measure given by Perelman $\cal L$-length functional.
\end{proof}

From this theorem we can state the following:

\begin{corollario}
   There exists a Wiener process for the Ricci flow of a two-dimensional Riemannian manifold.
\end{corollario}

From these results we can state the following conjecture:

\begin{congettura}
   There exists a Wiener process for the Ricci flow of a Riemannian manifold
\end{congettura}

This conjecture implies that one can modify Perelman $\cal L$-length functional, in a way to obtain the corrections to the heat equation for the metric, able to reproduce the full Ricci flow. These modifications should grant the existence of the corresponding Wiener integral. 





\end{document}